# DEVELOPMENT OF COGGING AT THE FERMILAB BOOSTER *

K. Seiya[#], S. Chaurize, C. Drennan, W. Pellico, A. K. Triplett, A. Waller, Fermilab, Batavia, IL 60510, USA

## Abstract

The development of magnetic cogging is part of the Fermilab Booster upgrade within the Proton Improvement Plan (PIP) [1]. The Booster is going to send 2.25E17 protons/hour which is almost double the present flux, 1.4E17 protons/hour to the Main Injector (MI) and Recycler (RR). The extraction kicker gap has to synchronize to the MI and RR injection bucket in order to avoid a beam loss at the rising edge of the extraction and injection kickers. Magnetic cogging is able to control the revolution frequency and the position of the gap using the magnetic field from dipole correctors while radial position feedback keeps the beam at the central orbit [2]. The new cogging is expected to reduce beam loss due to the orbit changes and reduce beam energy loss when the gap is created. The progress of the magnetic cogging system development is going to be discussed in this paper.

## CURRENT OPERATION AND GOAL FOR THE NEW COGGING

The Booster is a resonant circuit synchrotron with an operating frequency of 15Hz, which is synchronized to the 60Hz power line. Variations in the power line frequency and voltage result in deviation of the Booster cycle's length and starting point for acceleration. The magnetic field error changes the revolution frequency through the cycle which results in a change in position of the extraction kicker gap. The cogging process controls the position of the extraction kicker gap and synchronizes it to the desired MI or RR injection bucket.

The Booster accelerates the proton beam from 400MeV to 8GeV and extracts to the MI or RR. The beam is injected from the LINAC for 30 μsec with a 200MHz structure and is captured within 37.7MHz rf buckets adiabatically over 400 μsec. Since the Booster harmonic number is 84, the 84 bunches fill the Booster ring with 37.7MHz rf after the capture. The extraction kicker gap creation occurs at about 700 MeV, which is about 6 ms into the cycle for current operation. The new cogging method is going to create a gap as soon as the adiabatic capturing is finished. Creating a gap at 400MeV instead of 700MeV will reduce beam energy loss at the notching kicker location.

The current cogging process synchronizes the extraction kicker gap to the MI by changing the radial position of the beam during the cycle and creates unexpected beam loss due to the reduction of aperture. The new magnetic cogging process is going to change the dipole field using correctors and keeps the beam position on the central orbit.

The final bucket position has to be within +/- 1 target bucket before the Booster to MI phase lock process begins approximately 4 msec before beam extraction from the Booster.

## RF COUNTS AND MAGNETIC FIELDS

The cogging process makes the total number of rf counts through a cycle same as a reference cycle. One to twelve Booster batches are injected to the MI in one MI cycle and the first Booster cycle is used as a reference cycle. RF count through the reference cycle is captured and written to a memory table approximately every 10 μsec on the 1st Booster cycle. The relationship between frequency, magnetic field and radial position is written as

$$\frac{\Delta f_{rev}}{f_{rev}} = \frac{1}{\gamma^2}\frac{\Delta p}{p} - \frac{\Delta L}{L} = \frac{1}{\gamma^2}\frac{\Delta B}{B} - \frac{\Delta L}{L}. \quad (1)$$

where $f_{rev}$ is revolution frequency, $p$ is momentum, $B$ is dipole field and $L$ is circumference. When the Booster radial position feedback is regulated to a fixed orbit the relationship in eq. (1) becomes

$$\frac{\Delta f_{rev}}{f_{rev}} = \frac{1}{\gamma^2}\frac{\Delta B}{B}. \quad (2)$$

The magnetic cogging controls the revolution frequency by changing the magnetic fields. Figure 1 shows the block diagram of a control loop. The rf count difference (Δcount) between current cycle and the reference cycle is determined approximately every 10 μsec and this difference is integrated through the entire Booster cycle. A gain, G, is applied to the integrated count difference and the resulting signal is sent to the corrector power supplies. The total dipole field in the Booster is then the sum of the fields due to the dipole correctors and the main gradient magnets.

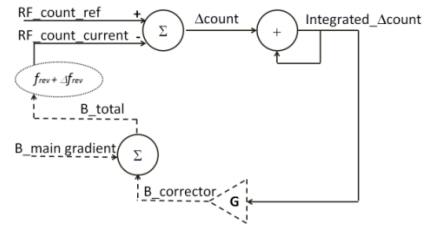

Figure 1: Block diagram of the cogging control loop.

*Work supported by Fermilab Research Alliance, LLC under Contract No. DE-AC02-07CH11359 with the United States Department of Energy.
#kiyomi@fnal.gov

# COGGING SYSTEM

The rf count is performed on a programmable VXI controller module. A digital to analog, DAC, output of the module produces a feedback signal for the dipole correctors, and two digital outputs of the module provide triggers to the notching kicker and the MI synchronization module as shown in Figure 2.

## The Cogging Programmable Controller

The controller uses 4 input signals, Bdot: beginning of acceleration, BRF: Booster RF signal, MIRF: MI RF signal, OAA: MI revolution marker and trigger event. The control process is as follows
1. Start counting BRF at Bdot.
2. Count the BRF cycles that occur within every MI revolution period (approx. 10 μsec).
3. Compare the count with the one from the reference cycle and take their difference.
4. Integrate the difference.
5. Multiply the integral by a gain from a gain table.
6. Send the product via the DAC output to the correctors.

The VXI controller module has 4 each, 14 bit DAC outputs. Figure 3 shows output signals from 3 of the DAC's; integrated count difference, the gain curve and the product of the integrated count difference times the gain, respectively from top to bottom. The integrated count difference multiplied by the gain curve is the cogging feedback signal. External buffers amplify the cogging feedback signal and distribute this signal around the Booster Gallery to the dipole corrector power supply controllers.

Bucket position delay and timing for notch creation can be changed via the accelerator control system, ACNET. Four 4096 x 16 bit memory buffers are available for diagnostics. This data can be plotted and saved as files using custom programs written in JAVA, running in ACNET.

## Corrector magnet

Booster has 24 periods. One period is made up of a 6 m long straight section, a defocusing magnet (D magnet), a 0.5 m mini straight section, a focusing magnet (F magnet), a 1.2 m short straight section, an F magnet, mini straight section and a D magnet. Twenty four correctors are located in long straight section between D magnets and 24 correctors are in short section between F magnets. The integrated field gradient of the dipole corrector is 0.009 T-m at 24.4A, and the slew rate is 3.24 T-m/s.

## Notcher

The Notcher is a horizontal kicker which creates a gap in the Booster beam and sends the beam kicked out into absorber. The kicker length is 70 nsec which is 2.5 buckets at the injection energy and 3.5 bucket at extraction. Three bunches are kicked out with the Notcher and the rest of the 81 bunches are sent to the MI and RR.

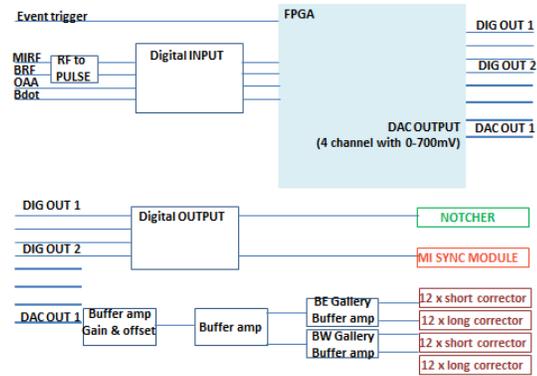

Figure 2: Cogging system layout.

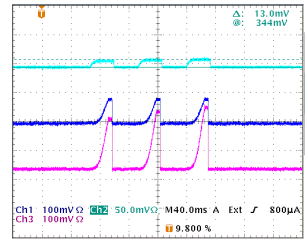

Figure 3: The DAC outputs, count difference, the gain curve and the product of the integrated count difference times the gain, respectively from top to bottom.

# FIELD ERROR MEASUREMENTS AND SIMULATIONS

## Field error and difference of counts

The variations of the gradient magnet field result in variations of revolution frequencies. Four sources of field error, with respect to the reference cycle, were assumed and used for simulations (Figure 4). The integration of the difference of rf counts were calculated for cases with 0.1% field error at injection, 0.1% field error at extraction, 100 μsec delay on cycle and 100 μsec error on cycle length. In reality, the error is expected to be combinations of these sources of error. The difference of rf counts were measured on 10 pulses and plotted on Figure 4.

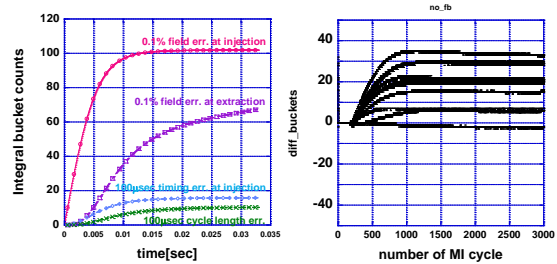

Figure 4: Count error on simulation (left) and count error from 10 pulse measurements (right).

*Simulation and measurements with feedback*

In order to optimize the feedback gain to make the final integrated difference in rf counts to be 0, measurements and simulations were performed using a fixed gain and a time varying gain curve. The number of rf buckets that the dipole correctors can move in 1 msec with 1A through each corrector magnet was calculated and plotted in Figure 5 (left). Since more current on the dipoles is required to move the same number of buckets at higher energy, a gain table (Figure5, right) was determined using the inverse of the results in Figure 5 (left). One rf count results in 0.2A through the correctors with the constant gain in both simulations and measurements. The gain curve values are shown in Figure 5 (right). Figure 6 shows results from 10 measurements where the final notch positions had a variation of +/- 2 buckets with constant gain and +/- 1 bucket using the gain curve. Figure 7 shows results from simulations which assume 0.1% field error at injection.

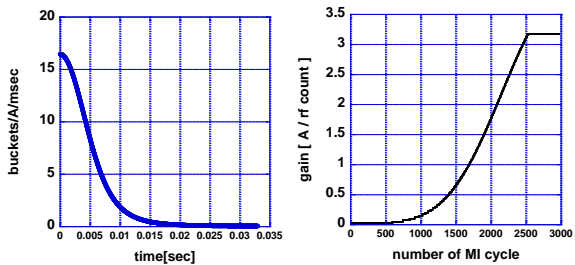

Figure 5: The bucket slippage in 1 msec with 1 A on 48 correctors. (left) A gain curve (right).

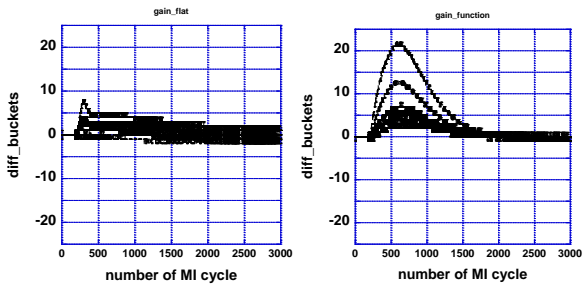

Figure 6: Measured rf counts with constant gain (left) and gain curve (right).

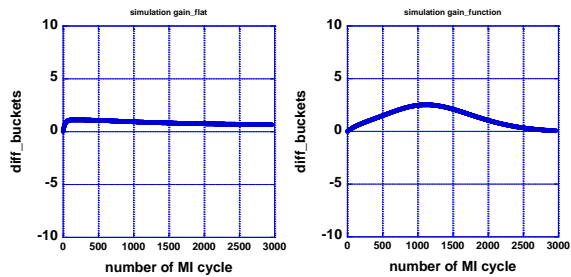

Figure 7: RF counts on simulations with constant gain (left) and gain curve (right).

## SYNCHRONIZATION TO THE MI

The extraction kicker gap is supposed to line up with a revolution maker (OAA) which represents the MI and RR desired injection bucket. Since the MI and RR stay at their injection energy, the period of OAA stays constant at approximately 10 μsec during the Booster cogging cycle. Even with the total rf counts in a cycle regulated to match the reference cycle, the time delay between OAA and Bdot is changing from cycle to cycle. The total number of bucket counts on a reference cycle is also changing from MI/RR cycle to cycle. Three delays determine the trigger time of the Notcher for a synchronous transfer to the MI or RR.

1. The delay between Bdot and OAA.
2. The delay between Bdot and OAA on the reference cycle.
3. The constant signal delay between the digital output of the cogging controller and the input to the Notcher firing system.

Four Booster batches were sent to the MI on machine study cycles. The mountain range plot, in Figure 8, shows that the 2nd to 4th batch were injected from the Booster to the MI every 66.6 msec. The resistive wall monitor was measured for 5.6 μsec on each trace and for 0.22 sec in vertical direction. The left plot in Figure 8 shows that the kicker gaps lined up with an edge of the batch with cogging. The right plot in Figure 8 shows the kicker gap in the middle of the batch without cogging.

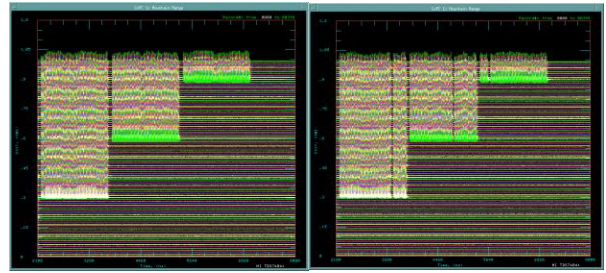

Figure 8: Mountain range signals in the MI.

## CONCLUSION

Magnetic cogging was able to control the position of the extraction gap using the dipole correctors. The magnetic cogging allows the extraction kicker gap to be created anytime in a cycle and reduces beam energy loss at the kicker location. Final bucket position was within +/- 1 target bucket with feedback as expected. When beam is sent to MI, the extraction kicker gap was synchronized to the desired MI bucket.

## REFERENCES

[1] W. Pellico et al., "FNAL - The Proton Improvement Plan (PIP) " ,THPME075, IPAC'14, to be published.
[2] K. Seiya et al., "Momentum Cogging at the Fermilab Booster", IPAC'12, New Orleans, 2012.